\documentclass[RNAAS]{aastex62}
\usepackage{amsmath}
\newcommand{\msun}{M$_{\odot}$}
\newcommand{\gdrat}{$\Delta_{\rm g/d}$}

\begin{document}

\title{Inferring the Gas-to-Dust Ratio in the Main Planet-Forming Region of Disks}

%%%\title{Inferring Disk Gas and Dust Depletion from Refractory Abundances}

%% Note that the corresponding author command and emails has to come
%% before everything else. Also place all the emails in the \email
%% command instead of using multiple \email calls.
\correspondingauthor{Adam S. Jermyn}
\email{adamjermyn@gmail.com}

\author[0000-0001-5048-9973]{Adam S. Jermyn}
\affiliation{Center for Computational Astrophysics, Flatiron Institute, New York, NY 10010, USA}

\author[0000-0003-0065-7267]{Mihkel Kama}
\affiliation{Department of Physics and Astronomy, University College London, Gower Street, London, WC1E 6BT, UK }
\affiliation{Tartu Observatory, Observatooriumi 1, Tõravere 61602, Tartu, Estonia}

\begin{abstract}
Measuring the amount of gas and dust in protoplanetary disks is a key challenge in planet formation studies. Here we provide a new set of dust depletion factors and relative mass surface densities of gas and dust for the innermost regions of a sample of protoplanetary disks. We do this by combining stellar theory with observed refractory element abundances in both disk hosts and open cluster stars. Our results are independent of, and complementary to, those obtained from spatially resolved disk observations.
\end{abstract}


\section{Introduction}

%Accretion disks can become depleted in gas, or dust, 

Frequently, the dust surface density in protoplanetary disks is measured by spatially resolving its thermal emission and assuming it is optically thin. This assumption is often valid, but not in the inner disk, up to tens of au from the star. This measurement is also insensitive to particles significantly larger than the emission wavelength (typically $0.1<\lambda<10\,$mm). The gas surface density is mainly estimated via one of two methods \citep{2018arXiv180709631B}: either observations of the rare but rotationally emissive H$_{2}$ isotopolog, HD \citep{2013Natur.493..644B, 2016ApJ...831..167M, 2017A&A...605A..69T, 2020A&A...634A..88K} or of optically thin $^{12}$CO isotopologs \citep{2016A&A...594A..85M, 2016ApJ...830...32W, 2019ApJ...883...98Z, 2022ApJ...925...49R}. Both of the above require ``upscaling'' number abundances to that of H$_{2}$ using assumed isotope ratios and elemental abundances, with uncertainties from chemical networks and volatile element depletion particularly in the case of CO. The gas-to-dust mass ratio (\gdrat) in disks is thus dependent on two quantities whose values are rather uncertain, especially in the inner disk.

Fortunately, for young stars with a mass ${\gtrsim}1.4\,$\msun\ (Herbig\,Ae/Be stars), a combination of slow internal mixing and high accretion rate leads to the photospheric elemental composition being dominated by recently accreted disk material. This makes the stellar spectrum a direct probe of the gas-to-dust mass ratio in the inner disk where the accretion originates. Indeed, the surface abundance of rock-forming elements (Fe, Mg, Si, etc.) in Herbig\,Ae/Be stars whose disks have a large dust-depleted inner region (a dust cavity extending outside the nominal dust sublimation front) is lower than the abundance in stars whose disks are ``full'' i.e., where the dust can freely accrete onto the star \citep{2015A&A...582L..10K}.

\section{Methods}

To accurately use the stellar surface abundances as a probe of the relative ratio of volatile gas to refractory dust in the inner disk, we need the stellar elemental composition for a sample of disk hosts, a stellar mixing model, and a reference composition to anchor the dust depletion factor to. 

To explain out approach we focus on iron.
To start with, we assume that the stars formed with an iron abundance comparable to that in a sample of non-accreting field stars, and that the bulk of each accreting star has this abundance ($X_{\rm Fe, bulk}$).
In each accreting star, some fraction $f_{\rm acc}$ of the photospheric material comes from the accretion stream and has composition $X_{\rm Fe,acc}$.
We calculate this using the CAMstars formalism~\citep{2018MNRAS.476.4418J}, which accounts for a variety of stellar mixing processes to match the accretion rate with the rate of mixing between the photosphere and the bulk of the star.
The observed photospheric abundance $X_{\rm Fe,obs}$ is then related to the bulk composition by
\begin{align}
    \label{eq:X}
    X_{\rm Fe,obs} = f_{\rm acc} X_{\rm Fe,acc} + (1-f_{\rm acc}) X_{\rm Fe, bulk}.
\end{align}
See section 2.4 of~\citet{2019ApJ...885..114K} for further discussion of this approach, which is related to those by~\citet{Turcotte_2002,1993ApJ...413..376T,1991ApJ...370..693C} and references therein.

We further assume that the accretion stream comes from the circumstellar disk, and that this disk formed with the same composition as the star.
As a refractory element, iron is locked in dust in the accretion stream, and so differences between $X_{\rm Fe,acc}$ and $X_{\rm Fe,bulk}$ reflect a change in the gas-to-dust ratio due to e.g. dust traps or gas accretion onto planets.
We thus define the \emph{dust enhancement} factor
\begin{align}
    \delta \equiv \frac{X_{\rm Fe,acc}}{X_{\rm Fe,bulk}}.
    \label{eq:d}
\end{align}
This is what we aim to infer.

We use the same sample of stars as in~\citet{2019ApJ...885..114K}, see appendix A1 for details.
In brief, abundances were taken from \citet{2011MNRAS.413.1132F}, \citet{2012MNRAS.422.2072F}, \citet{2016A&A...592A..83K}, and \citet{2017MNRAS.466..613M}.

We use abundances for elements Fe, Mg, Si, Ca, Sc, and Ti in our inference procedure, treating each as we treated iron above and using uniform $\delta$ and $f_{\rm acc}$ across elements. 
We then employ the Bayesian Multinest sampling algorithm~\citep{doi:10.1111/j.1365-2966.2007.12353.x,doi:10.1111/j.1365-2966.2009.14548.x,1306.2144} to obtain posterior probability distributions on $\log \delta$ for each star individually.
To allow for the possibility of either a very strong dust depletion or a dust enhancement up to \gdrat\,$=1$, we set a uniform prior on $\log \delta$ over $[-3,2]$ and simultaneously infer $f_{\rm acc}$, using a gaussian prior distribution calculated with the mean and variance from our CAMstars calculation.
Our likelihood is defined as a Gaussian in $\log X$, fitting the observed iron abundance for each star using the field sample and equations~\eqref{eq:X} and~\eqref{eq:d}.
Summary statistics from this inference are reported in Table~\ref{tab:table}. We additionally report the more familiar gas-to-dust mass ratio: \gdrat\,$=100\times\delta^{-1}$, where $100$ is the standard value, \gdrat\,${>}100$ characterises dust depletion as in a cavity or gap in the dust disk, and ${<}100$ means enhancement, perhaps from a large influx of pebbles to the innermost disk region.

\section{Results}

The gas-to-dust mass ratios presented in Table\,\ref{tab:table} reveal both dust-enriched (\gdrat\,${<}100$) and -depleted inner disks (${>}100$). For this sample, all systems with dust-enriched inner disks are so-called Group\,II or ``full'' dust disks where pebbles are expected to drift into the inner disk relatively unobstructed, while the dust-depleted inner disks all belong to the ``transitional'' category, which are usually disks with a very prominent dust trap (ring) and an inner cavity. Two of the dust-depleted inner disks, while thought to be ``full'' on large scales, display dust ring structures on $\approx$ sub-au scales \citep[see][for discussion and references]{2015A&A...582L..10K}.

As an example, models of the disk around HD\,163296 have favoured \gdrat\,${<}100$ for the inner regions \citep{2016MNRAS.461..385B}, consistent with our inferred value of $81$. Similarly favourable comparisons can be made for disk models and our measurements in the case of HD\,100546. A thorough cross-analysis is however outside the scope of this Research Note, where we aim to provide our measured \gdrat\ values as an inner boundary condition for any analysis of the gas and dust surface density profile in the disks. In particular, our results characterise the disk material that is accreting onto the star, implying it is characteristic of the innermost few astronomical units where most planets are thought to complete their assembly.

\begin{deluxetable}{lrllllll}[t]
    \tablecolumns{8}
    \tablecaption{Summary statistics for the inferred dust enhancement factor $\delta$ and photospheric accretion fraction $f_{\rm acc}$ are shown for the stars in our accreting sample. Values for each are given as median, $-1\sigma$, and $+1\sigma$. We additionally report the gas-to-dust ratio using the median $\log \delta$,  calculated as \gdrat\,$=100\times\delta^{-1}$.\label{tab:table}}
    \vspace{0.5em}
    \tablehead{\colhead{Star} & \colhead{\gdrat} & \colhead{$\log \delta$} & \colhead{$-1\sigma$} & \colhead{$+1\sigma$} & \colhead{$\log f_{\rm acc}$} & \colhead{$-1\sigma$}
    & \colhead{$+1\sigma$}}
    % & Disk type}
    \startdata
\multicolumn{8}{c}{Dust-depleted innermost disk}\\
 HD169142  &       5129 &      -1.71 &       -2.62 &       -0.72 &      -0.35 &       -0.46 &       -0.25 \\
 HD141569  &       4673 &      -1.67 &       -2.56 &       -0.75 &      -0.47 &       -0.79 &       -0.31 \\
 T Ori     &       2890 &      -1.46 &       -2.51 &       -0.50 &      -0.41 &       -0.57 &       -0.27 \\
 HD144432  &       1443 &      -1.16 &       -2.41 &       -0.24 &      -0.61 &       -0.84 &       -0.41 \\
 HD142666  &      987 &      -0.99 &       -2.33 &       -0.25 &      -0.41 &       -0.55 &       -0.23 \\
 HD100546  &      301 &      -0.48 &       -0.56 &       -0.41 &      -0.01 &       -0.01 &       -0.00 \\
 HD68695   &      210 &      -0.32 &       -0.40 &       -0.25 &      -0.01 &       -0.01 &       -0.00 \\
 HD245185  &      194 &      -0.29 &       -0.37 &       -0.20 &      -0.01 &       -0.01 &       -0.00 \\
 HD 278937 &      194 &      -0.29 &       -0.35 &       -0.22 &      -0.01 &       -0.01 &       -0.00 \\
 HD101412  &      182 &      -0.26 &       -0.34 &       -0.18 &      -0.01 &       -0.01 &       -0.00 \\
 HD139614  &      169 &      -0.23 &       -0.30 &       -0.16 &      -0.01 &       -0.01 &       -0.00 \\
 HD179218  &      161 &      -0.21 &       -0.28 &       -0.13 &      -0.01 &       -0.01 &       -0.00 \\
 \hline
 \multicolumn{8}{c}{Dust-enriched innermost disk}\\
 %\hline
 HD36112   &     97 &       0.01 &       -0.06 &        0.08 &      -0.01 &       -0.01 &       -0.00 \\
 HD244604  &     86 &       0.06 &       -1.70 &        0.49 &      -0.92 &       -1.32 &       -0.49 \\
 HD31648   &     84 &       0.07 &        0.01 &        0.14 &      -0.01 &       -0.01 &       -0.00 \\
 HD163296  &     81 &       0.09 &       -0.28 &        0.28 &      -0.52 &       -0.79 &       -0.25 \\
    \enddata
\end{deluxetable}

\section{Data Availability}

The CAMstars software instrument is available on \href{https://github.com/adamjermyn/CAMstars}{GitHub}. The analysis in this work was performed with the \texttt{individual\_open\_cluster\_Fe.py} script, also in that repository.

\acknowledgments

The Flatiron Institute is supported by the Simons Foundation.

\bibliography{refs}
\bibliographystyle{aasjournal}

\end{document}